\documentclass[fleqn,10pt]{wlscirep}

\title{Active Brownian particles and run-and-tumble particles separate inside a maze}

\author[1,2,*]{Maryam Khatami}
\author[2]{Katrin Wolff}
\author[2]{Oliver Pohl}
\author[1,3]{Mohammad Reza Ejtehadi}
\author[2]{Holger Stark}
\affil[1]{Department of Physics, Sharif University of Technology, P.O. Box 11155-9161, Tehran, Iran}
\affil[2]{Institut für Theoretische Physik, Technische Universität Berlin, Hardenbergstrasse 36, 10623 Berlin, Germany}
\affil[3]{School of Nano-Science, Institute for Research in Fundamental Sciences (IPM), P. O. Box 19395-5531, Tehran, Iran}

\affil[*]{m.khatami.m@gmail.com}

\begin{abstract}
A diverse range of natural and artificial self-propelled particles are known and are used nowadays. Among them, active Brownian particles (ABPs) and run-and-tumble particles (RTPs) are two important classes. We numerically study non-interacting ABPs and RTPs strongly confined to different maze geometries in two dimensions. We demonstrate that by means of geometrical confinement alone, ABPs are separable from RTPs. By investigating \textit{Matryoshka}-like mazes with nested shells, we show that a circular maze has the best filtration efficiency. Results on the mean first-passage time reveal that ABPs escape faster from the center of the maze, while RTPs reach the center from the rim more easily. According to our simulations and a rate theory, which we developed, ABPs in steady state accumulate in the outermost region of the \textit{Matryoshka}-like mazes, while RTPs occupy all locations within the maze with nearly equal probability. These results suggest a novel technique for separating different types of self-propelled particles by designing appropriate confining geometries without using chemical or biological agents.
\end{abstract}

\begin{document}
\flushbottom
\maketitle
\thispagestyle{empty}

\noindent Active particle systems have attracted a lot of attention during the last decades \cite{Ramaswamy2010a,Marchetti2013,Zottl2016}. Much effort has been devoted to study and mimic the motion of motile bacteria, molecular motors, and other natural low-Reynolds-number swimmers \cite{Ajdari1997,Polin2009,Bennett2013,Najafi2004,Leoni2009,Golestanian2005,Howse2007,Schmitt2013,Kummel2013b,Ebrahimian2015,Alizadehrad2015,Adhyapak2015}. With the significant advance of micro- and nanotechnology, different powerful artificial micro- and nanoswimmers have been fabricated. Examples include Janus particles, catalytic micro-jets and magnetic nano-propellers \cite{Ghosh2009,Jiang2010,Volpe2011,Restrepo-Perez2014}.

With respect to their swimming strategies, self-propelled particles are categorized into two main classes. On the one hand, one considers run-and-tumble particles (RTPs) such as motile \textit{E. Coli} bacteria and \textit{Chamydomonas algae} \cite{Polin2009}. Basically, these microswimmers move on straight lines, \textit{runs}, with constant speed $v_0$ and reorient randomly, \textit{tumble}, with rate $ \alpha $. On the other hand, active Brownian particles (ABPs) refer to the second class of self-propelled particles with spherical Janus particles and non-tumbling mutant strains of \textit{E. Coli} \cite{Tavaddod2011} as examples. This type of active particles moves with constant speed $v_0$ but their direction changes gradually due to their rotational diffusivity $ D_r $. Despite their different moving strategies, RTPs and ABPs behave similarly on large length and time scales; like passive Brownian particles (PBP) they exhibit translational  diffusion. More precisely, a pure ABP and a pure RTP diffuse with the same translational diffusion constant, if $ (d-1) D_r = \alpha $, where $ d $ is spatial dimension \cite{Cates2013}. Furthermore, they sediment almost similarly in a gravitational field \cite{Tailleur2009,Palacci2010,Enculescu2011,Solon2015} or become trapped in harmonic potentials \cite{Nash2010,Pototsky2012,Hennes2014,Solon2015}.

Several experimental and theoretical studies have demonstrated that after an encounter with a wall microswimmers slide along the wall for some time, before they leave it  \cite{Galajda2007,Volpe2011,Denissenko2012,Kantsler2013,Schaar2015}. This peculiar scattering behavior can cleverly be employed to rectify the motion of microswimmers in micro- and nanochambers \cite{Galajda2007,Wan2008,Denissenko2012,Kantsler2013,Koumakis2014,Das2015}, to trap these particles \cite{Kaiser2013,Chepizhko2013,Reichhardt2014,Takagi2014,Sipos2015,Spagnolie2015}, and to extract mechanical work from their random motion \cite{Angelani2009,DiLeonardo2010,Sokolov2010,Koumakis2013,Kaiser2014}. Furthermore, since microswimmers have to reorient to leave a surface, both RTPs and ABPs strongly accumulate at bounding walls in the regime of strong confinement, where the persistence length of straight runs is much larger than the box dimension \cite{Rothschild1963,Berke2008,Angelani2009,Volpe2011,Fily2014,Vladescu2014,Gompper2015}. However, the density peak observed at the wall disappears by increasing the size of the surrounding box, \emph{i.e.}, when leaving this regime \cite{Vladescu2014}.

Despite of their similar properties in bulk and confinement on large scales, ABPs and RTPs interact differently with obstacles on small scales \cite{Brown2015}. These differences can have a substantial impact. For example, wild-type and aflagellated mutants of \textit{Salmonella} or other microbes infect host cells with different efficiencies \cite{Achouri2015}. So, it is of crucial importance to separate different types of active particles or microswimmers from each other. It has already been demonstrated that active particles can be separated from passive colloids by using funnel-shaped barriers \cite{Galajda2007} or the tendency of active particles to accumulate in corners \cite{Kaiser2012, Restrepo-Perez2014, Guidobaldi2014}. Different types of active particles are also separable from each other according to their chiralities and linear or angular speeds \cite{Mijalkov2013}. However, a powerful tool or strategy to separate ABPs and RTPs from each other is missing.

\begin{figure}[t]
	\centering
	\includegraphics[width=0.6\linewidth]{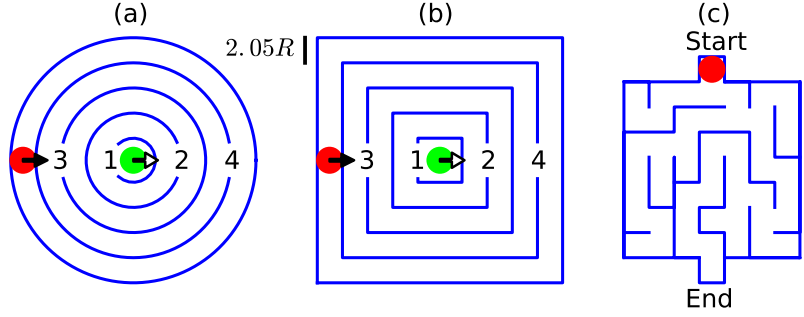}
	\caption{\textbf{Circular (a), square (b), and non-regular maze (c) used in Brownian dynamics simulations.} The disks indicate the initial particle positions, where simulations start. In (a) and (b) green refers to the \textit{outwards} case and red to the \textit{inwards} case, respectively. Arrows indicate the initial velocity direction and numbers show the opening index.}
	\label{fig:Fig1_all-mazes}
\end{figure}

In this article, we present such a strategy by investigating active particles in suitable geometries analogous to mazes, which are illustrated in Fig.\ \ref{fig:Fig1_all-mazes}. Mazes or labyrinths are interesting complex geometries and solving them, \emph{i.e.}, proceeding from the entrance to the exit, has been a challenging problem for many centuries. For example, using chemotaxis, active particles are able to solve mazes and find the shortest path towards the exit \cite{Nakagaki2000, Lagzi2010, Reynolds2010}. Here, we consider non-interacting ABPs and RTPs inside the different two-dimensional mazes of Fig.\ \ref{fig:Fig1_all-mazes}, but without any chemical field. Using computer simulations to determine mean first-passage times and stationary probability distributions, we demonstrate that ABPs are faster than RTPs in moving from the center to the rim of circular and square mazes, while RTPs reach the center more easily. This suggests the nested regular mazes of Fig.\ \ref{fig:Fig1_all-mazes}(a) and (b) as excellent tools for separating active particles from each other, with an advantage for the circular shape as we will demonstrate.

\section*{Materials and Methods}

We consider a single self-propelled particle that is either an ABP or RTP. We place it at a specific starting point inside the maze and monitor its dynamics until it reaches the target point using two-dimensional Brownian dynamics simulations. To obtain the full statistics for an ensemble of non-interacting active particles, we perform many simulation runs to generate different realizations for the same initial condition.

\subsection*{Particle Dynamics} 

The position vectors of both model particles evolve according to the same over-damped dynamics
\begin{equation}
	\dot{\vec{r}} = v_0 \hat{e} + \vec{F}_{w} .
	\label{eq:r-dynamics}
\end{equation}
Here, the overdot indicates time derivative, $v_0$ is the particle's self-propulsion speed, which is kept constant during the simulations, and its intrinsic direction of motion is given by the unit vector $\hat{e} = \cos (\theta) \hat{x} + \sin (\theta) \hat{y} $, where $ \theta $ is the orientation angle with respect to an arbitrarily chosen $x$ axis. The particle interacts with walls by a simple repulsive force (see e.g. \cite{Fily2014,Fily2015}) implemented such that the total particle velocity along the normal of the wall is zero at contact:
\begin{eqnarray}
	\vec{F}_w(\vec{r}, \theta) = 
	\left \{
	\begin{array}{lr}
		-v_0 (\hat{e} \cdot \hat{n}  ) \,
		\hat{n} \qquad \qquad & \, |\Delta \vec{r}_w| \leq R \\
		{}\qquad  & {} \\
		0 \qquad \qquad & 
		\mathrm{otherwise} 
	\end{array}\right.
	\label{}
\end{eqnarray}
The unit vector $\hat{n}$ indicates the local normal, $ R $ is the particle radius, which sets the length scale of our problem, and $ \Delta \vec{r}_w $ denotes the closest distance between particle center and nearest point on the wall. Since $\vec{F}_w$ removes the normal component of the self-propulsion velocity, active particles slide along the walls, which reproduces experimental observations \cite{Galajda2007,Volpe2011,Denissenko2012,Kantsler2013}. In the following, we neglect translational diffusion for both types of particles ($ D_t = 0 $). More details about the numerical method are provided in the supplementary material.

Most importantly, ABP and RTP differ in the dynamics governing their orientation angle. An ABP changes its self-propulsion direction continuously due to rotational diffusion. Thus,
\begin{equation}
	\dot{\theta}(t) = \sqrt{2 D_r}\eta(t),
	\label{eq:theta-ABP-dynamics}
\end{equation}
where $ D_r $ is the particle's rotational diffusion constant and $ \eta $ is unbiased ($ < \eta (t) > = 0 $) and uncorrelated ($ <\eta(t) \eta(t')> = \delta(t-t')$) white noise. In contrast, an RTP changes its direction abruptly with rate $\alpha$ at each time step,
\begin{equation}
	\dot{\theta}(t) = \sum_i \Delta \theta_i \delta(t-T_i).
	\label{eq:theta-RTP-dynamics}
\end{equation}
The sum runs over all tumble events that stochastically happen at times $T_i$. The tumble angle $ \Delta \theta_i $ is selected with equal probability from the range between $ 0$ and $ 2\pi $. 

At long times the motion of ABPs and RTPs becomes diffusive in unconfined space. The respective diffusion constants agree if the equivalence condition $(d-1) D_r = \alpha $ (with $d$ the spatial dimension) is fulfilled \cite{Cates2013}. We will use this relation to compare the results for ABPs and RTPs to each other. In two dimensions, $ \tau_r = 1/D_r$ is the persistence time of an ABP, on which the particle orientation decorrelates. For an RTP the equivalent time $\tau_r = 1/\alpha$ is the mean run time between two tumble events. To characterize the motion of both particle types, we will use the \emph{persistence number} 

\begin{equation}
P_r = \frac{v_0 \tau_r}{2R},
\end{equation}
which compares the length of straight runs to the particle diameter, and vary it in the range $ P_r \in [1.5, 75.0] $.

To investigate the single-particle behavior of ABPs and RTPs, we propose three geometries analogous to mazes; a circular, square, and non-regular (random) maze. Each maze is characterized as follows. 

\subsection*{Types of Mazes}

\subsubsection*{Circular maze}

Figure\ \ref{fig:Fig1_all-mazes}(a) shows a picture of the circular maze. It consists of five concentric circular shells (walls) nested inside each other, similar to a Russian \textit{Matryoshka} doll. The inner-shell radius is $ r_1 = 1.8 R $ and the other ones obey $ r_{i} - r_{i-1} = 2.05 R$ (for $ i = 2, 3, 4, 5 $). So the width of each circular annulus (zone) is a little larger than the diameter of the particle. Using these values, the particle effectively moves on one-dimensional circular paths with varying speed depending on particle orientation $\hat{e}$. Except of the outer shell, all other shells have an opening to let the particle transfer between adjacent zones. All these openings have the same width equal to $3 R$ and are opposite to each other for neighboring shells.

We study the ability of both active-particle types to solve the maze in two cases. In the first case, we initially place the particle in the central zone of the maze and wait until it reaches  the outer zones (this case is briefly called \textit{outwards}). In the second case, we initially place the particle in the outermost annulus and wait until it reaches the central zone of the maze (\textit{inwards}). In both cases, the initial particle orientation is along $ + \hat{x}$. We measure the mean first-passage time (MFPT) for the particle to pass through each opening [as numbered in Fig.\ \ref{fig:Fig1_all-mazes}(a)]. The numerical simulations run over 200 realizations for the outwards and between 30-200 for the inwards case.

\subsubsection*{Square maze}

We investigate the effect of wall curvature by considering a square maze analogous to the circular one. As shown in Fig. \ref{fig:Fig1_all-mazes}(b), the square maze consists of five concentric square shells nested inside each other. Their edge lengths equal the diameter of the corresponding circle in the circular maze. Initial conditions are similar to the circular maze. The results for the MFPTs are obtained by averaging over 200 initializations for both the outwards and inwards case.

\subsubsection*{Non-regular maze}

We consider a random (non-regular) maze to find out how the absence of regularity of the previous geometries affects the particle behavior. The non-regular maze studied here [Fig. \ref{fig:Fig1_all-mazes}(c)] is a branched maze with several dead ends, one entrance, one exit, and a single solution. It is constructed in such a way that its lane width equals the zone width of the \textit{Matryoshka}-like mazes described before. The self-propelled particle is initially placed at the point of the maze marked Start in Fig. \ref{fig:Fig1_all-mazes}(c) with random velocity direction and the MFPT is measured for the particle to reach the End point. Results are obtained by averaging over 200 initializations.

\begin{figure}[t]
	\centering
	\includegraphics[width=0.6\linewidth]{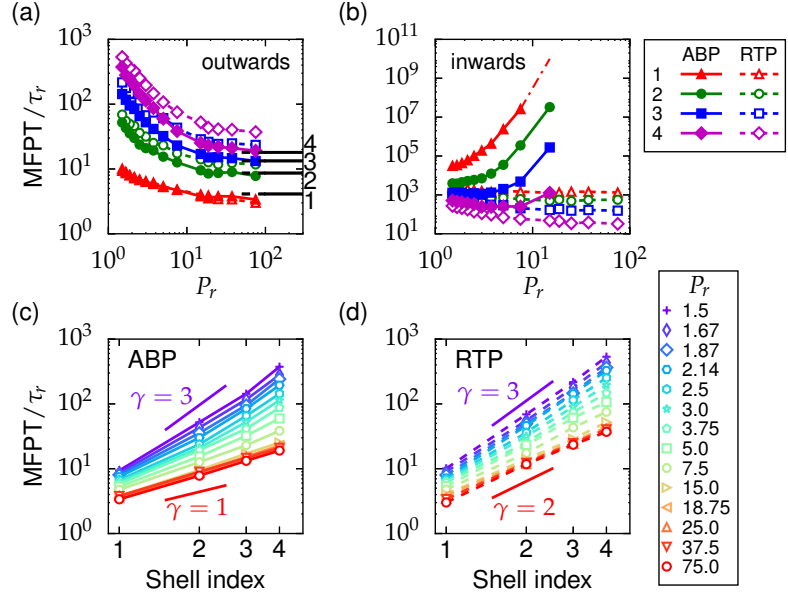}
	\caption{\textbf{MFPT of an ABP and an RTP for the circular maze in the outwards (a, c, d) and inwards case (b)}. Symbols refer to simulation data. They are connected by lines to guide the eyes. Error bars are smaller than symbols and do not exceed 6\% of the mean value. (a), (b): MFPT versus $P_r$ for reaching the opening with index $i$ as indicated by the legend. In (a) the black horizontal lines show theoretical predictions obtained from Eq. (\ref{eq:ABP_equ}). The red dashed-dotted line in (b) is an extrapolation of the simulation data to larger $P_r$. (c),(d) MFPT versus shell index $i$ for different $P_r$ for an ABP (c) and an RTP (d). Characteristic slopes $\gamma$ in the double-logarithmic plot are indicated. The data in (c) and (d) are the same as in (a).}
	\label{fig:Fig2_mfpt_circular}
\end{figure}

\section*{Results}

\subsection*{Circular maze (outwards)}

We find that the MFPT not only changes systematically with persistence number $P_r$ but it also depends on the type of the self-propelled particle. Figure \ref{fig:Fig2_mfpt_circular}(a) shows the MFPT of both particle types for the outwards initialization. For both types of particles, the MFPT decreases with $P_r$. However, the ABP (solid symbols) always reaches the openings in the confining walls faster than the RTP (open symbols). Both observations can be understood by carefully analyzing how the self-propelled particles move along the walls. At large persistence number $P_r$ (strong confinement regime), the ABP performs a random walk along convex boundaries. While its mean orientation is aligned with the local boundary normal, the momentary orientation fluctuates about the normal and thereby leads to short drift motions \cite{Fily2014}. By sliding along the walls, the ABP readily finds the wall openings and escapes through them successively (see supplementary movie 1). In contrast, the RTP detaches from the wall during a tumble event where it reorients abruptly (see supplementary movie 2). This strategy seems to be less efficient for finding the openings, since the RTP does not use the guiding walls as much as the ABP does. Thus, the MFPTs are larger in this case. By decreasing $P_r$, both particles more often detach from the walls and the exit times increase accordingly.

\begin{figure} [t]
	\centering
	\includegraphics[width=0.75\linewidth]{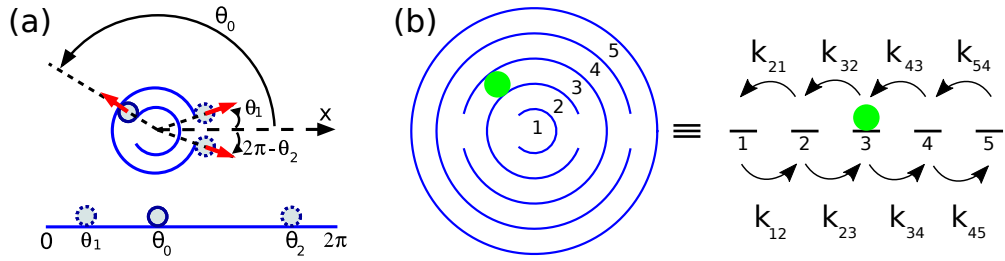}
	\caption{(a) Theory for MFPT for large $P_r$: Parameters used in deriving Eq. (\ref{eq:ABP_equ}) are indicated. (b) Linear rate theory: The circular maze is mapped on a one-dimensional lattice. Each zone of the maze is equivalent to a site on this lattice. $k_{mn}$ denotes the transition rate to jump from zone/site $m$ to zone/site $n$.}
	\label{fig:Fig3_two-theory}
\end{figure}

In case of an ABP, a simple random-walk model describes the asymptotic behavior of the MFPT in the limit of large persistence number $P_r$. In this case the particle orientation is mostly aligned with the local boundary normal and the orientation angle $\theta$ also describes the location of the ABP in the annulus [see Fig.\ \ref{fig:Fig3_two-theory}(a)]. We can now calculate the MFPT to go from one wall opening to the next one and thereby to pass to the next zone in the maze. The total MFPT is then the sum over all openings reached. The orientation angle just changes by rotational diffusion. So, in order to escape through an opening, the orientation angle has to diffuse from $\theta_0$ (the position of the previous opening) either to $\theta_1$ or $\theta_2$ [see Fig. \ref{fig:Fig3_two-theory}(a)]. This is equivalent to  the first-passage problem of a particle diffusing in one dimension with two absorbing boundaries at $\theta_1$ and $\theta_2$. Using the standard formalism \cite{muthukumar2016polymer,gardiner2004handbook}, which we summarize in the supplementary material, the MFPT for a particle initially at $\theta_0$ for reaching one of the absorbing boundaries can be calculated:
\begin{equation}
	T = \frac{(\theta_0 - \theta_1) (\theta_2 - \theta_0)}{2 D_r}
	\label{eq:ABP_equ}
\end{equation}
To calculate the MFPT for reaching opening $i$, we choose $\theta_0 = \pi$. Furthermore, the angles $\theta_1$ and $\theta_2$ depend on the shell index, which we summarize in the supplementary material. As the black horizontal lines in Fig.\ \ref{fig:Fig2_mfpt_circular}(a) demonstrate, we obtain good agreement with the simulation results at large $P_r$ (see supplementary material for a quantitative comparison). At small $P_r$, the intrinsic direction $\hat{e}$ is no longer aligned with the local boundary normal and the described theory is not applicable any more. In this regime, the self-propelled particle frequently detaches from the wall due to the more pronounced rotational diffusion. As a result, the mean exit time becomes larger than the asymptotic expression of Eq. (\ref{eq:ABP_equ}). This behavior is confirmed in Fig.\ \ref{fig:Fig2_mfpt_circular}(a).

We also investigated how the MFPT increases from one shell to the other by plotting the MFPT versus the opening or shell index $i$. As shown in Figs.\ \ref{fig:Fig2_mfpt_circular}(c) and (d), for increasing persistence number the slope of the lines in the double-logarithmic plot declines from $\gamma = 3$ to $\gamma = 1$ for ABP and from $\gamma = 3$ to $\gamma = 2$ for RTP. As explained in the previous paragraph, at large $P_r$ the MFPT of an ABP to proceed from one opening to the next one is purely governed by rotational diffusion and thus the same for all openings. Therefore, the total MFPT increases linearly in $i$, which gives the slope $\gamma = 1$. At small $P_r$ the intrinsic direction $\hat{e}$ is no longer perpendicular to the wall. Its rotational diffusion is so fast that the ABP performs an effective translational diffusion along the annulus with radius $ r_i \propto i$. So, the MFPT to reach and escape from opening $i$ scales with $r_i ^2 \propto i^2$ and the total MFPT grows proportional to $i^3$, which gives $\gamma = 3$. The same explanation also holds for RTPs at small $P_r$. 
At large $P_r$ the distances between successive reorientation events are also large (see supplementary movie 2). So a few ballistic steps are sufficient to reach opening $i$ and we can roughly estimate the mean exit time as proportional to $r_i \propto i$. Thus, the total MFPT grows like $i^2$, which gives $\gamma =2$.

We also determined the probability density function for the first-passage times (FPT) needed to reach opening $i$. Examples are presented in Fig. S2 of the supplementary material. They all reveal an exponential decay with an existing MFPT.

\subsection*{Circular maze (inwards)}
The most interesting result arises by initializing the self-propelled particles from the outermost zone of the maze [inwards case in Fig. \ref{fig:Fig1_all-mazes}(a)]. We find that the ABP enters the inner zones of the maze much less readily than the RTP. As demonstrated in Fig.\ \ref{fig:Fig2_mfpt_circular}(b), the MFPT for the ABP is several orders of magnitude larger than for the RTP and diverges with increasing $P_r$. This is because at large $P_r$, the velocity vector of the ABP always points radially outward and rotational diffusion typically only allows for small changes in its direction. Thus, when arriving at an opening, it is very unlikely that the ABP fully reverses its orientation $\hat{e}$ to enter the opening. The probability to enter the opening decreases drastically with increasing $P_r$ since the ABP is faster and therefore spends less time close to the opening. Thus, the MFPT ultimately diverges. In contrast, an RTP always has a finite probability to reverse its direction and thereby to enter the openings towards the inner parts of the maze. We also find a non-monotonic behavior in the MFPT for an ABP entering the $4^{th}$ opening. At small $P_r$ or small $v_0$ the ABP hardly moves towards the opening. Thus, the time to reach it and thereby the MFPT increases again with decreasing $P_r \propto v_0$. A similar non-monotonic behavior is also reported in recent results on optimal search strategies for an ABP looking for a target at the center of a confining circular domain \cite{Wang2016}.

\begin{figure} 
	\centering
	\includegraphics[width=0.61\linewidth]{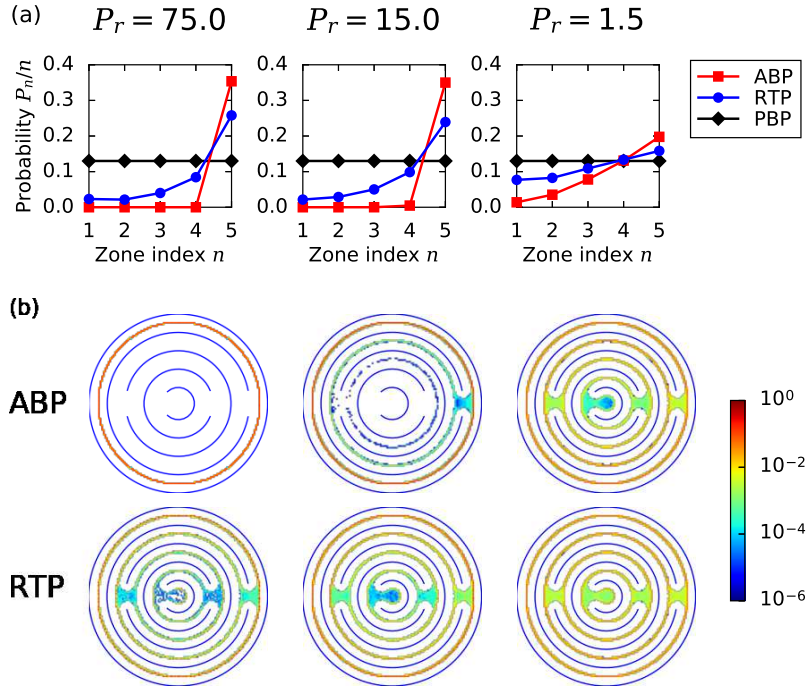}
	\caption{\textbf{Stationary distribution of active and passive particles inside the circular maze}. 
	a) Probability $p_n/n$ for occupying zone $n$ for different $P_r$. PBP refers to passive Brownian particle.
	b) Probability density within the maze color-coded by a logarithmic scale covering six orders of magnitude.}
	\label{fig:Fig4_densitymap_circular}
\end{figure}
%
When we let our simulations run for a sufficiently long time, both in the outwards and the inwards case, the particle distribution reaches the same steady state and the particles occupy zone $n$ in the circular maze with probability $p_n$. In Fig.\ \ref{fig:Fig4_densitymap_circular}(a), we plot $p_n / n$ for different $P_r$, where the zone index $n$ is proportional to the zone radius. Clearly, this probability is constant for PBPs that do not have any internal driving. They spread over the whole maze with uniform density just by thermal diffusion. However, at large $P_r$, ABPs only accumulate in the outermost zone (index 5), while the inner zones remain unoccupied. This is in agreement with the diverging MFPTs plotted in Fig.\ \ref{fig:Fig2_mfpt_circular}(b) for the inwards case. Only for $P_r \approx 1$ do ABPs also reach the center of the maze. In contrast, RTPs with their ability to perform large reorientations also settle in the inner zones even at large $P_r$. All these findings are nicely illustrated in Fig.\ \ref{fig:Fig4_densitymap_circular}(b), where we directly plot the color-coded probability density in the circular maze. So by using this maze, RTPs can be separated from ABPs very well under confinement.

When generating the steady state distributions, one has to be sure that the system has relaxed to the steady state starting from an initial condition. To estimate the relevant relaxation times, we developed a linear rate theory. The probabilities $p_n(t)$ for occupying zone $n$ are elements of the state vector $|\psi(t)> = |p_1, p_2, p_3, p_4, p_5>$, which evolves in time according to
\begin{equation}
	\frac{d}{dt} |\psi (t)> = - {\bf{Q}} |\psi (t)> \, .
	\label{eq:rate-theory}
\end{equation}
The transition matrix $\bf{Q}$ contains the transition or hoping rates $k_{mn}$ between neighboring zones as introduced in Fig.\ \ref{fig:Fig3_two-theory}(b). The detailed form of $\bf{Q}$ is given in the supplementary material. For different $P_r$, we determined the rates $k_{mn}$ from simulations of the outwards case, which corresponds to the initial condition $|\psi(t)> = |(1, 0, 0, 0, 0)>$. The eigenvalues of $\bf{Q}$ are the relaxation rates of characteristic relaxation modes. One zero eigenvalue belongs to the steady state, which we discussed in Fig.\ \ref{fig:Fig4_densitymap_circular}. The inverse of the next smallest eigenvalue gives the largest relaxation time $\tau_{\mathrm{relax}}$, one generally has to wait before steady state is reached. We plot $\tau_{\mathrm{relax}}$ versus $P_r$ in Fig.\ S4 in the supplementary material. 

\subsection*{Square maze}

\begin{figure} 
	\centering
	\includegraphics[width=0.7\linewidth]{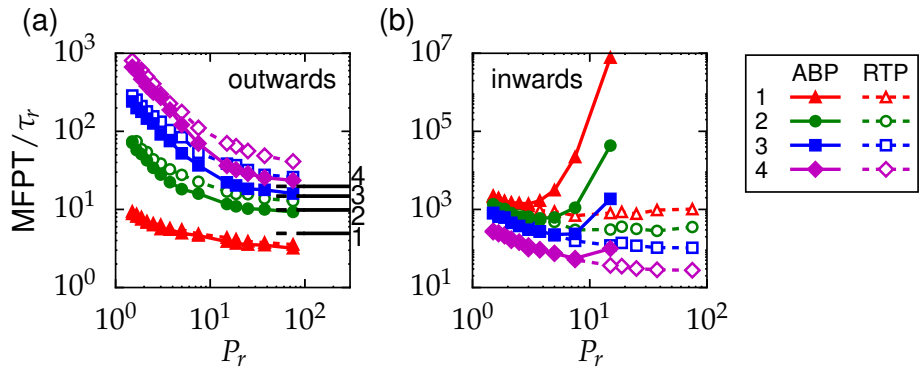}
	\caption{\textbf{MFPT versus persistence number $P_r$ for the square maze in the outwards (a) and inwards case (b)}. Symbols refer to simulation data. They are connected by lines to guide the eyes. Error bars are smaller than symbols. In (a) the black horizontal lines show theoretical predictions obtained from Eq. (\ref{eq:ABP_equ}) using $\theta_0 = \pi$, $\theta_1 = 0$, and $\theta_2 = 2 \pi$. The adjacent numbers refer to the opening index.}
	\label{fig:Fig5_mfpt_square}
\end{figure}

To find out how wall curvature affects particle separation, we examine the square maze of Fig. \ref{fig:Fig1_all-mazes}(b) with square walls nested inside each other. The simulation results for the MFPTs are similar to the circular maze [compare Fig. \ref{fig:Fig5_mfpt_square} to Fig.\ \ref{fig:Fig2_mfpt_circular}(a),(b)]. In the outwards case the ABP escapes from the inner zones faster than the RTP [see Fig. \ref{fig:Fig5_mfpt_square}(a)], while in the inwards case the RTP enters the inner zones more easily with MFPTs close to the values of the circular maze [see Fig. \ref{fig:Fig5_mfpt_square}(b) and supplementary movie 5]. However, a careful inspection of the inwards case reveals an important difference between the two mazes; the strong increase of the MFPT with increasing $P_r$ for an ABP is more pronounced in the circular maze [see Fig.\ \ref{fig:Fig2_mfpt_circular} (b)]. Thus, the MFPTs for ABPs differ significantly in both mazes by several orders of magnitude for increasing $P_r$. In addition, while in the circular maze the MFPTs of ABP and RTP at small $P_r$ differ from each other, especially for the innermost opening [see Fig.\ \ref{fig:Fig2_mfpt_circular} (b)], they are nearly identical in the square maze. We consider these differences to be due to wall curvature. As argued before, ABPs tend to stay at walls with positive curvature for large $P_r$, while they easily leave walls with negative curvature \cite{Fily2014,Fily2015}. Therefore, an ABP in the circular maze needs to make a full $180^{\circ}$ turn close to an opening in order to enter this opening, which is a rare event. In contrast, in the square maze an ABP can slide on both sides of the straight walls. Therefore, by rotational diffusion it can face the wall with the opening and then just slide to the opening to enter it (see supplementary movie 6). This scenario obviously takes less time compared to the circular maze and the MFPTs of the square maze are smaller.

\begin{figure}
	\centering
	\includegraphics[width=0.6\linewidth]{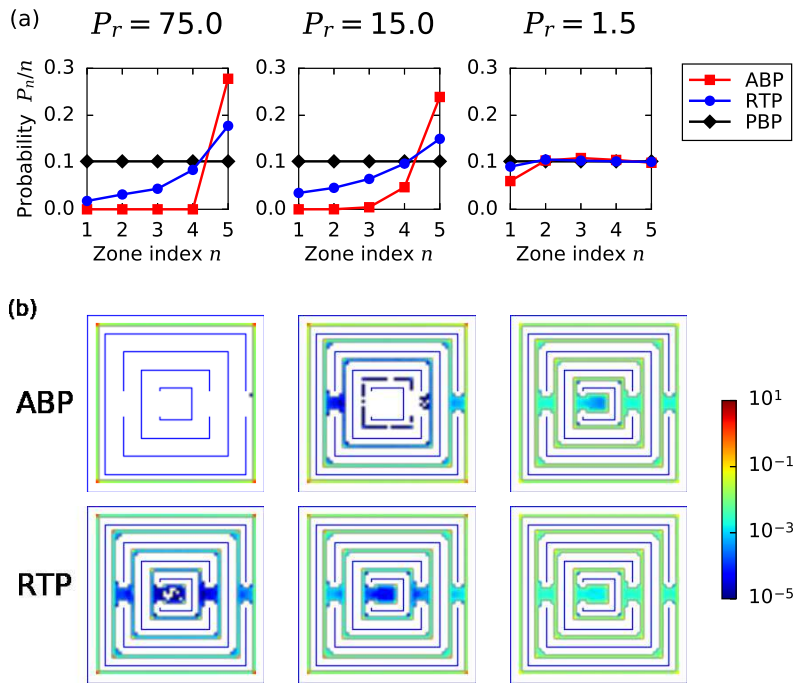}
	\caption{\textbf{Stationary distribution of active and passive particles inside the square maze}. 
	a) Probability $p_n/n$ for occupying zone $n$ for different $P_r$. PBP refers to passive Brownian particles.
	b) Probability density within the maze color-coded by a logarithmic scale covering six orders of magnitude.}
	\label{fig:Fig6_densitymap_square}
\end{figure}
%

The steady-state distributions of the square maze plotted in Fig.\ \ref{fig:Fig6_densitymap_square} show similar behavior compared to the circular maze (see Fig.\ \ref{fig:Fig4_densitymap_circular}). However, we realize that the differences between the ABP, RTP, and PBP are not as strong. In particular, at $P_r=1.5$ all particle types now show a uniform distribution throughout the square maze. In addition, the active particles strongly accumulate in the corners of the square maze, especially at large $P_r$, which is in line with previous experimental observations on both ABPs \cite{Fily2014,Guidobaldi2014,Restrepo-Perez2014,Kaiser2012} and RTPs \cite{Angelani2009,DiLeonardo2010,Sokolov2010}
and other theoretical works with different geometries\cite{Ray2014,Takagi2014,Sipos2015}. Finally, we have also formulated the rate theory based on Eq.\ (\ref{eq:rate-theory}). The largest relaxation time, necessary for identifying the steady state, is plotted in Fig. S3 of the supplementary material.

\subsection*{Non-regular maze}

\begin{figure}
	\centering
	\includegraphics[width=0.7\linewidth]{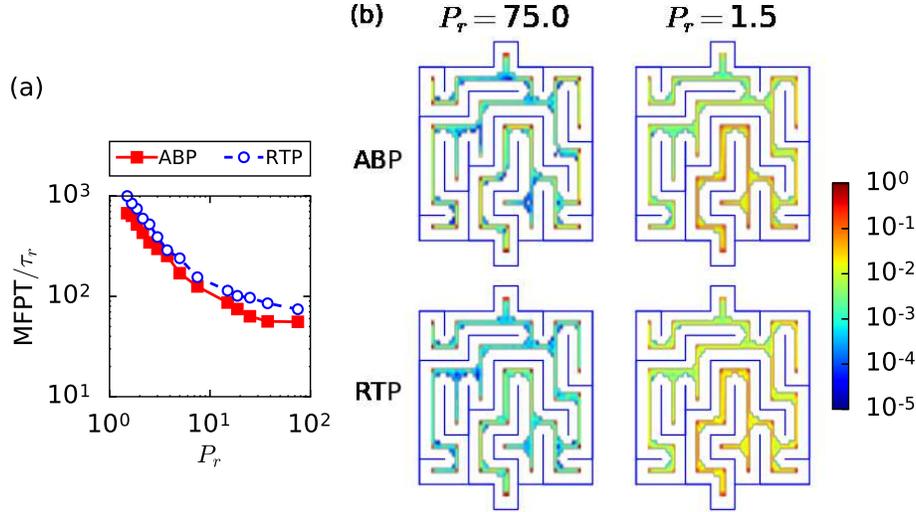}
	\caption{\textbf{ABP and RTP inside the non-regular maze.}
	(a) MFPT versus $P_r$. Error bars have the same size as the symbols. 
	(b) Stationary probability distributions of an ABP (first row) and an RTP (second row). The color bar shows the logarithmic scale for probability covering 5 orders of magnitude.}
	\label{fig:Fig7_non-regular}
\end{figure}
%
%

Finally, we investigated the non-regular maze of Fig. \ref{fig:Fig1_all-mazes}(c). In contrast to the nested mazes, no significant difference between the MFPTs of an ABP and an RTP are observable in Fig. \ref{fig:Fig7_non-regular} (a). This is due to two features: first of all, the pattern is irregular and second, nested structures as in the regular mazes are absent. In particular, the second feature generated the difference between both types of active particle. For example, whenever an ABP, starting from the outermost zone of the circular maze, enters the $4^{th}$ opening at large $P_r$, it is more probable for it to return to the outermost zone rather than to enter the $3^{rd}$ opening, since for most of the time its orientation points radially outwards. Therefore, the nested structure of Matryoshka-like mazes drastically decreases the probability for the ABP to enter the inner regions of the maze and amplifies the difference between the two active-particle types. This feature is missing in the non-regular maze.

We also determined the spatial probability distribution of both active-particle types in steady state [see Fig. \ref{fig:Fig7_non-regular}(b)]. It does not show any significant difference between ABPs and RTPs neither at low nor at high persistence number $P_r$. However, corner accumulation is still observable in this maze and it is more pronounced at large $P_r$. Interestingly, at $P_r = 1.5$ the particle density increases from the top (starting point) to the bottom (end point), while for large $P_r$ the distribution is more uniform throughout the maze. Finally, we tested whether the size of the maze affected these results by performing simulations with a maze twice as large but with the same structure. Again, no difference was observed between both active-particle types (results not shown).

\section*{Discussion}

In this article we have explored the behavior of two important and widespread types of active particles inside different mazes. Namely, non-interacting active Brownian particles and run-and-tumble particles strongly confined in Matryoshka-like mazes show large differences both in their mean first-passage times and their stationary probability distributions. In particular, run-and-tumble particles enter the maze and easily move towards the center, while active Brownian particles escape faster towards the rim. In steady state, persistent ABPs accumulate in the outermost region of the maze, while RTPs spread more uniformly in all regions. This peculiar phenomenon is more pronounced in mazes with curved boundaries. Our work suggests Matryoshka-like mazes as excellent devices to separate different types of active particles from each other with potential technological and biomedical applications. In contrast, a random maze is not able to distinguish between both active-particle types.

In experiments, regular mazes can be fabricated using microfabrication techniques \cite{Koumakis2013,Galajda2007,Restrepo-Perez2014} and used to efficiently separate different types of self-propelled particles without any biological or chemical agent. Our work also has biomedical relevance for separating bacteria from other microorganisms or passive objects such as blood cells. In future work it will be interesting to study the collective behavior of active particles in these strongly confining complex structures \cite{Grossman2008a,Seyed-Allaei2016} and to explore their potential for optimal search strategies \cite{Wang2016,Rupprecht2016,Reichhardt2014}.

\bibliography{myreferences1}

\section*{Acknowledgements}
We would like to acknowledge invaluable discussions with Seyed Hamid Seyed-allaei. We also thank Andreas Z{\"{o}}ttl for fruitful discussions at an early stage of the work, Ramin Golestanian and Julien Tailleur for useful comments on the project, and Pietro Cicuta for bringing reference \cite{Achouri2015} to our attention. This work was supported in part by DFG through the research training group GRK 1558 and the priority program SPP 1726 (grant number STA352/11), Iran National Science Foundation (grant number 93031724), and Iran's National Elites Foundation.






\end{document}